\documentclass[11pt,prd,letterpaper,amsmath,amssymb,final,nofootinbib
,unsortedaddress,superscriptaddress
]{revtex4-1}
\usepackage{geometry}[letter,11pt]
\usepackage{graphicx}
\usepackage[utf8]{inputenc}
\usepackage{xcolor}
\usepackage{times}
\usepackage{rotating}
\usepackage{geometry}
\usepackage{graphicx}
\usepackage{dcolumn}
\usepackage{bbm}
\usepackage{textcomp}
\usepackage{color}
\usepackage{gensymb}
\usepackage{wasysym}
\usepackage{amssymb}
\usepackage{enumerate}
\usepackage{geometry}
\usepackage{setspace}
\usepackage{comment}
\usepackage{lineno}
\usepackage[utf8]{inputenc}	
\usepackage{mathrsfs}
\usepackage[T1]{fontenc}
\usepackage{amssymb,amsmath}
\usepackage{graphicx, amsthm}
\usepackage{booktabs}
\usepackage{isotope}
\usepackage[graphicx]{realboxes}
\usepackage{xspace}
\usepackage{enumitem}
\setlist[enumerate]{itemsep=0mm}

\newcommand \theia{\textsc{Theia} }
\newcommand \eos{\textsc{Eos} }

\begin{document}
\title{\textsc{Eos}: a demonstrator of hybrid optical detector technology}
\author{T.~Anderson}\affiliation{\psu}
\author{E.~Anderssen}\affiliation{\lbnl}
\author{M.~Askins}\affiliation{\lbnl}\affiliation{\ucb}
\author{A.~J.~Bacon}\affiliation{\penn}
\author{Z.~Bagdasarian}\affiliation{\lbnl}\affiliation{\ucb}
\author{A.~Baldoni}\affiliation{\psu}
\author{N.~Barros}\affiliation{\fcul}\affiliation{ \lip}
\author{L.~Bartoszek}\affiliation{\bart}
\author{M.~Bergevin}\affiliation{\llnl}
\author{A.~Bernstein}\affiliation{\llnl}
\author{E.~Blucher}\affiliation{\chic}
\author{J.~Boissevain}\affiliation{\bart}
\author{R.~Bonventre}\affiliation{\lbnl}
\author{D.~Brown}\affiliation{\lbnl}
\author{E.~J.~Callaghan}\affiliation{\lbnl}\affiliation{\ucb}
\author{D.~F.~Cowen}\affiliation{\psu}
\author{S.~Dazeley}\affiliation{\llnl}
\author{M.~Diwan}\affiliation{\bnl}
\author{M.~Duce}\affiliation{\lbnl}\affiliation{\gt}
\author{D.~Fleming}\affiliation{\css}
\author{K.~Frankiewicz}\affiliation{\bu}
\author{D.M.~Gooding}\affiliation{\bu}
\author{C.~Grant}\affiliation{\bu}
\author{J.~Juechter}\affiliation{\rut}
\author{T.~Kaptanoglu}\affiliation{\lbnl}\affiliation{\ucb}
\author{T.~Kim}\affiliation{\rut}
\author{J.R.~Klein}\affiliation{\penn}
\author{C.~Kraus}\affiliation{\snolab}\affiliation{\laur}
\author{T.~Kroupov\'a}\affiliation{\penn}
\author{B.~Land}\affiliation{\lbnl}\affiliation{\ucb}\affiliation{\penn}
\author{L.~Lebanowski}\affiliation{\lbnl}\affiliation{\ucb}\affiliation{\penn}
\author{V.~Lozza}\affiliation{\fcul}\affiliation{ \lip}
\author{A.~Marino}\affiliation{\boul}
\author{A.~Mastbaum}\affiliation{\rut}
\author{C.~Mauger}\affiliation{\penn}
\author{G.~Mayers}\affiliation{\penn}
\author{J.~Minock}\affiliation{\rut}
\author{S.~Naugle}\affiliation{\penn}
\author{M.~Newcomer}\affiliation{\penn}
\author{A.~Nikolica}\affiliation{\penn}
\author{G.~D.~Orebi~Gann}\affiliation{\lbnl}\affiliation{\ucb}
\author{L.~Pickard}\affiliation{\lbnl}\affiliation{\ucb}\affiliation{\ucd}
\author{L.~Ren}\affiliation{\boul}
\author{A.~Rincon}\affiliation{\lbnl}\affiliation{\ucb}
\author{N.~Rowe}\affiliation{\lbnl}\affiliation{\ucb}
\author{J.~Saba}\affiliation{\lbnl}
\author{S.~Schoppmann}\affiliation{\lbnl}\affiliation{\ucb}\affiliation{\mainz}
\author{J.~Sensenig}\affiliation{\penn}
\author{M.~Smiley}\affiliation{\lbnl}\affiliation{\ucb}
\author{H.~Song}\affiliation{\bu}
\author{H.~Steiger}\affiliation{\mainz}\affiliation{\mun}
\author{R.~Svoboda}\affiliation{\ucd}
\author{E.~Tiras}\affiliation{\erc}\affiliation{\iow}
\author{W.~H.~To}\affiliation{\css}
\author{W.~H.~Trzaska}\affiliation{\jyv}
\author{R.~Van~Berg}\affiliation{\penn}\affiliation{\bart}
\author{V.~Veeraraghavan}\affiliation{\iowa}
\author{J.~Wallig}\affiliation{\lbnl}
\author{G.~Wendel}\affiliation{\psu}
\author{M.~Wetstein}\affiliation{\iowa}
\author{M.~Wurm}\affiliation{\mainz}
\author{G.~Yang}\affiliation{\lbnl}\affiliation{\ucb}
\author{M.~Yeh}\affiliation{\bnl}
\author{E.D.~Zimmerman}\affiliation{\boul}

\newcommand{\ucb}{Physics Department, University of California at Berkeley, Berkeley, CA 94720-7300
}
\newcommand{\lbnl}{
Lawrence Berkeley National Laboratory, 1 Cyclotron Road, Berkeley, CA 94720-8153, USA
}
\newcommand{\penn}{Department of Physics and Astronomy, University of Pennsylvania, Philadelphia, PA 19104-6396
}\newcommand{\fcul}{Universidade de Lisboa, Faculdade de Ci{\^e}ncias (FCUL), Departamento de F{\'i}sica, Campo Grande, Edifício C8, 1749-016 Lisboa, Portugal
}\newcommand{\lip}{Laborat{\'o}rio de Instrumenta{}{\c c}{\~a}o e F{\'i}sica Experimental de Part{\'i}culas (LIP), Av. Prof. Gama Pinto, 2, 1649-003, Lisboa, Portugal
}\newcommand{\chic}{
The Enrico Fermi Institute and Department of Physics, The University of Chicago, Chicago, IL 60637, USA
}\newcommand{\bnl}{
Brookhaven National Laboratory, Upton, New York 11973, USA
}\newcommand{\uh}{
University of Hawai‘i at Manoa, Honolulu, Hawai‘i 96822, USA
}\newcommand{\iowa}{
Department of Physics and Astronomy, Iowa State University, Ames, IA 50011, USA
}\newcommand{\jyv}{
Department of Physics, University of Jyv{\"a}skyl{\"a}, Finland
}\newcommand{\ucd}{
University of California, Davis, 1 Shields Avenue, Davis, CA 95616, USA
}\newcommand{\bu}{
Boston University, Department of Physics, Boston, MA 02215, USA
}\newcommand{\mainz}{
Johannes Gutenberg-Universit{\"a}t Mainz, 55099 Mainz, Germany
}\newcommand{\ham}{
Institut f{\"u}r Experimentalphysik, Universit{\"a}t Hamburg, 22761 Hamburg, Germany
}\newcommand{\alb}{
University of Alberta, Department of Physics, 4-181 CCIS, Edmonton, AB T6G 2E1, Canada
}\newcommand{\pnnl}{
Pacific Northwest National Laboratory, Richland, WA 99352, USA
}\newcommand{\laur}{
Laurentian University, Department of Physics, 935 Ramsey Lake Road, Sudbury, ON P3E 2C6, Canada
}\newcommand{\lsu}{
Department of Physics and Astronomy, Louisiana State University, Baton Rouge, LA 70803
}\newcommand{\tub}{
Kepler Center for Astro and Particle Physics, Universit{\"a}t T{\"u}bingen, 72076 T{\"u}bingen, Germany
}\newcommand{\sheff}{
University of Sheffield, Physics \& Astronomy, Western Bank, Sheffield S10 2TN, UK
}\newcommand{\qu}{
Queen's University, Department of Physics, Engineering Physics \& Astronomy, Kingston, ON K7L 3N6, Canada
}\newcommand{\snolab}{
SNOLAB, Creighton Mine 9, 1039 Regional Road 24, Sudbury, ON P3Y 1N2, Canada
}\newcommand{\rut}{
Department of Physics and Astronomy, Rutgers, The State University of New Jersey, 136 Frelinghuysen Road, Piscataway, NJ 08854-8019 USA
}\newcommand{\temp}{
Department of Physics, Temple University, Philadelphia, PA, USA
}\newcommand{\ucla}{
University of California Los Angeles, Department of Physics \& Astronomy, 475 Portola Plaza, Los Angeles, CA 90095-1547, USA
}
\newcommand{\tri}{
SISSA/INFN, Via Bonomea 265, I-34136 Trieste, Italy
}\newcommand{\kav}{
Kavli IPMU (WPI), University of Tokyo, 5-1-5 Kashiwanoha, 277-8583 Kashiwa, Japan
}\newcommand{\kor}{
Center for Underground Physics, Institute for Basic Science, Daejeon 34126, Korea
}\newcommand{\uci}{
University of California, Irvine, Department of Physics, CA 92697, Irvine, USA
}\newcommand{\sbu}{
State University of New York at Stony Brook, Department of Physics and Astronomy, Stony Brook, New York, USA
}\newcommand{\tsing}{
Department of Engineering Physics, Tsinghua University, Beijing 100084, China
}\newcommand{\corn}{
Cornell University, Ithaca, NY, USA
}\newcommand{\boul}{
University of Colorado at Boulder, Department of Physics, Boulder, Colorado, USA
}\newcommand{\dres}{
Institut f{\"u}r Kern und Teilchenphysik, TU Dresden, Zellescher Weg 19, 01069, Dresden, Germany
}
\newcommand{\mun}{Physics Department, Technische Universit{\"a}t M{\"u}nchen, 85748 Garching, Germany
}
\newcommand{\mitnew}{
Massachusetts Institute of Technology, Department of Physics and Laboratory for Nuclear Science, 77 Massachusetts Ave Cambridge, MA 02139, USA
}
\newcommand{\kings}{King’s College London, Department of Physics, Strand Building, Strand, London WC2R 2LS, UK}
\newcommand{\llnl}{
Lawrence Livermore National Laboratory, Livermore, CA 94550, USA
}
\newcommand{\fnal}{
Fermi National Accelerator Laboratory, Batavia, IL 60510, USA
}
\newcommand{\erc}{Department of Physics, Erciyes University, 38030, Kayseri, Turkey
}
\newcommand{\iow}{Department of Physics and Astronomy, The University of Iowa, Iowa City, Iowa, USA}
\newcommand{\psu}{Pennsylvania State University, University Park, PA 16802, USA}

\newcommand{\heid}{Ruprecht-Karls-Universitat Heidelberg, Im
Neuenheimer Feld 227, Heidelberg, Germany}

\newcommand{\bart}{Bartoszek Engineering, Aurora, IL 60506, USA}

\newcommand{\ucbne}{Nuclear Engineering Department, University of California at Berkeley, Berkeley, CA 94720-7300
}

\newcommand{\css}{California State University, Stanislaus, Department of Physics, Turlock, CA 95382, USA}

\newcommand{\gt}{Georgia Institute of Technology, Nuclear and Radiological Engineering, Atlanta, GA, 30313, USA}

\begin{abstract}
\eos is a technology demonstrator, designed to explore the capabilities of hybrid event detection technology, leveraging both Cherenkov and scintillation light simultaneously.  
With a fiducial mass of four tons, 
\eos is designed to operate in a high-precision regime, with sufficient size to utilize time-of-flight information for full event reconstruction,  flexibility to demonstrate a range of cutting edge technologies, and simplicity of design to facilitate potential future deployment at alternative sites.  Results from \eos 
can inform the design of future neutrino detectors for both fundamental physics and nonproliferation applications.
\end{abstract}

\maketitle 

\section{Introduction}

Neutrinos can offer insights into both fundamental physics and practical applications, such as probing the fusion reactions that power our Sun, studying the explosion mechanism of distant supernovae, monitoring  nuclear activity, and even explaining the origin of our matter-dominated Universe.  Low-energy neutrino detection sensitivity is limited by the small interaction cross section, and the presence of background events such as naturally occurring radioactivity, and cosmogenic muons and their spallation products.  Improvements in event identification and reconstruction of event properties can have a significant impact on the ability to distinguish background from signal and, therefore, serve to increase the physics reach of an experiment.

Decades of paradigm-defining experiments demonstrate the efficacy and impact of both Cherenkov and scintillation techniques for neutrino detection~\cite{SNO,SuperK,IMB,kamland,borexino,MiniBooNE,nova,T2K}.   
The ``hybrid'' technique for  neutrino detection leverages both Cherenkov and scintillation light simultaneously, resulting in event imaging capabilities that would be transformative for background rejection, 
offering world-leading sensitivity to probe a broad range of physical and application-driven phenomena.  The impact of this technology to a rich program of physics, including neutrinoless double beta decay, supernova and solar neutrinos, geoneutrinos, and long baseline neutrinos, is explored in papers such as Refs.~\cite{Aberle:2013jba,Bonventre:2018hyd,Askins:2019oqj,TheiaAntinu,Land:2020oiz,TheiaDSNB,Elagin:2016zgp}. Additionally, hybrid detectors are being investigated for far-field detection of antineutrinos for the purpose of nuclear nonproliferation~\cite{WATCHMAN:2015lcq,Bat:2021jyq}.

Many avenues are being explored to realise hybrid event detection.  Novel scintillators can help facilitate detection of a ``clean'' Cherenkov signal by modifying the Cherenkov-to-scintillation ratio or adjusting the scintillation time profile~\cite{Caravaca:2016fjg,CHESS2020,Biller:2020uoi,Guo:2017nnr}.  Photon detectors are being developed that improve light collection and time precision, with enhanced configurations that even offer spectral sensitivity~\cite{Lyashenko:2019tdj,osti_1564252,Klein:2022tqr,Kaptanoglu:2017jxo,Kaptanoglu:2019gtg}. New readout techniques seek to allow pulse counting across a large dynamic range from one to many photoelectrons. 

Hybrid detection has been employed successfully in experiments looking at high energy interactions -- LSND pioneered the technique by looking at Cherenkov light in a diluted liquid scintillator to reconstruct electron tracks in the energy range of about 45 MeV~\cite{Patterson:2009ki} and the MiniBooNE reconstruction utilized both the Cherenkov and scintillation light produced from high energy ($\sim$1 GeV) neutrino interactions in mineral oil~\cite{Reeder:1993ff}. Borexino has achieved the first directional sensitivity in a large-scale scintillator detector with a fast, high light yield scintillator by using the first detected photons in each event. The collaboration was able to demonstrate statistical evidence for solar neutrino directionality for sub-MeV neutrinos, integrated across the low-energy dataset~\cite{BOREXINO:2021efb}.  SNO+ has achieved the first event-by-event directional reconstruction of few-MeV scale events.  This was achieved with the detector half-filled with linear alkyl benzene (LAB) with a low admixture of 2,5-Diphenyloxazole (PPO)~\cite{snoplus_direc}: a lower-yield, slower scintillator than is ultimately planned for this detector~\cite{Anderson:2020xxb}.

On the bench-top scale, successful detection of Cherenkov light from fast, highly scintillating media has been achieved~\cite{CHESS2016,CHESS2020,Kaptanoglu:2019gtg,Gruszko:2018gzr,Kaptanoglu:2021prv,mainz} 
and complemented with an extensive characterization program on a national and international scale. 
These measurements inform large-scale Monte Carlo (MC) simulation models, which are used to predict performance in future kton-scale detectors. 

While bench-top measurements have been, and will continue to be used to measure microphysical properties of novel scintillators and photon detection technology,  demonstrations of event reconstruction and the resulting background rejection capabilities are still largely simulation driven.  A data-driven demonstration of the event imaging capabilities of hybrid detector technology, with the ability to probe how that performance changes with a changing detector configuration, is a critical step to realizing an optimized, dedicated large detector with a maximally broad physics program.

The goal of \eos is to provide a testbed for studying the impact of detector configuration on performance.  This four-ton (fiducial) detector offers sufficient size and a high level of instrumentation for full event reconstruction using photon time-of-flight information.  Full containment of a range of low-energy events allows detailed event-level characterization, complemented with flexibility to adapt for multiple target materials and photon detection options.
\eos will also allow benchmarking of existing MC models on a scale that will significantly improve confidence in scaling performance predictions to kton-- and many-kton-scale detectors. 

Complementary to \eos is an international program of novel scintillator development and deployment.
Ton-scale deployments of novel scintillators, such as water based liquid scintillator (WbLS)~\cite{YehWbLS}, are planned or ongoing. Four hundred kg of WbLS will be deployed in the Booster Neutrino Beam at ANNIE~\cite{sandi}, in order to demonstrate high-energy neutrino event reconstruction. A ton-scale deployment is ongoing at Brookhaven National Laboratory (BNL), and the performance and stability is being probed with cosmogenic muons~\cite{wbls_recent}. A 30-ton vessel is also planned at BNL, which will be the first large-scale (above 10-ton) deployment of WbLS.
The BNL 30-Ton WbLS (30TBNL) demonstrator is a multi-year development effort aiming to construct a fully operational testbed with the goal of making a precise measurement of optical properties of WbLS over long distances, demonstrating material compatibility with detector components, and additional characterizations, as a next-step towards the deployment of a kton-scale detector.
\eos, together with this suite of detectors, will inform the design of next-generation neutrino experiments.


\section{Detector goals}
The primary goal of \eos is to demonstrate the performance capabilities of hybrid detector technology, and to test MC-model predictions in order to validate performance predictions for large-scale next-generation experiments.  This includes a data-driven demonstration of full event reconstruction, including event position, direction, and energy. It will also have the flexibility to probe the dependence of event resolution on critical detector configuration parameters such as the light yield and time profile of the deployed scintillator, the photocoverage, photon detector time resolution, and the addition of spectral sorting.  A comparison of these results in data-to-model predictions will allow a precise validation of the MC models, which will support previous and future predictions for large-scale detector performance in next-generation experiments such as \theia~\cite{Askins:2019oqj}.

This will be achieved by deployment of radioactive sources to fully probe low-energy event reconstruction capabilities, and use of cosmogenic muons for high-energy event reconstruction and tracking.
The goals will be to:
\begin{enumerate}
    \item Demonstrate and quantify improved vertex reconstruction compared to water Cherenkov detectors;
    \item Demonstrate and quantify direction reconstruction for low-energy events by selecting the Cherenkov signature against the scintillation background;
    \item Demonstrate muon track reconstruction;
    \item Compare data to model predictions to validate event reconstruction capabilities;
    \item Explore particle identification capabilities resulting from both the species-dependent time profile and the Cherenkov-to-scintillation light ratio;
    \item Probe performance in data and model predictions for a range of different detector configurations: differing scintillator mixtures, photon detector response, and the addition of spectral sorting.
\end{enumerate}

\eos can provide an important testbed to the community for testing alternative target media, photon detectors, and readout technology and methodology, and assessing the impact of these novel developments on detector performance.  

In addition \eos is designed to preserve the option of future deployments near a reactor, or at a particle test beam such as at CERN, FNAL, or the Spallation Neutron Source (SNS) at Oak Ridge National Laboratory (ORNL)~\cite{SNS,SNS2,CERN}.  These would allow more advanced demonstrations of particle and event identification leveraging the unique hybrid signature.

\section{\eos Detector design}

\begin{figure}[!h]
    \begin{center}
    \includegraphics[width=0.95\textwidth]{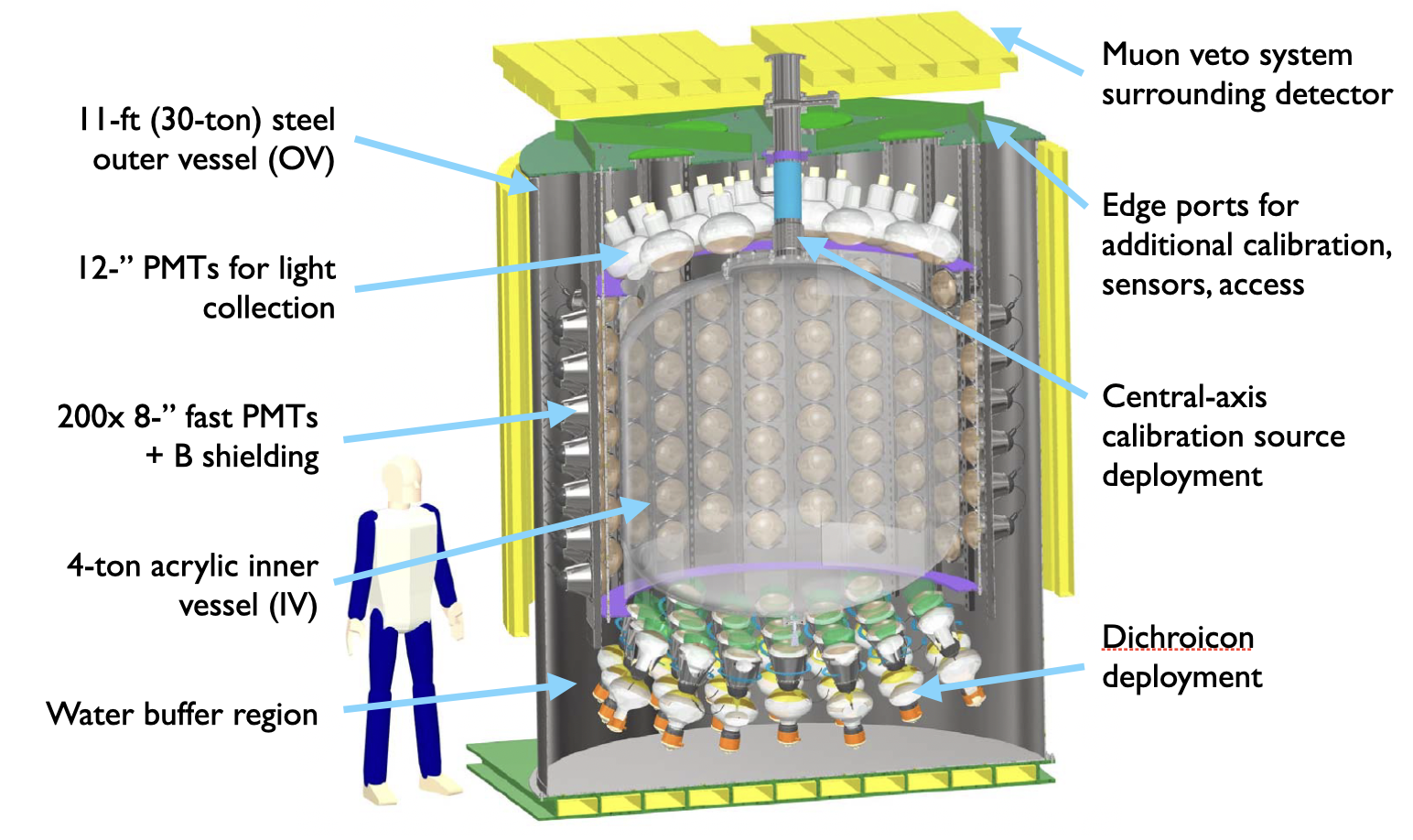}        
    \end{center}
    \caption{Conceptual design of the Eos detector, showing the different detector components.}
    \label{f:eos}
\end{figure}

\eos will consist of a 4-ton acrylic inner vessel (IV), which can house either a water or scintillator target.  The IV will be viewed by approximately 250 photomultiplier tubes (PMTs) with varying properties to facilitate an understanding of the dependence of detector performance on the photon detector configuration.  On the lower PMT array, a number of ``dichroicons'' will be installed, which achieve Cherenkov/scintillation separation via spectral sorting~\cite{Kaptanoglu:2018sus, Kaptanoglu:2019gtg}.  The detector assembly will be housed in a stainless steel outer vessel (OV), with a source deployment mechanism to allow deployment of low-energy radioactive and optical sources down a central axis.  Figure~\ref{f:eos} shows the conceptual design of the detector.

Two hundred 8-inch R14688-100 PMTs from Hamamatsu with better than 1-ns time precision will facilitate Cherenkov photon identification.  Two dozen 12-inch R11780 PMTs~\cite{Brack:2012ig} from Hamamatsu will offer enhanced light collection on the top array.  A combination of 5-inch green-sensitive R6594 PMTs
and high-quantum efficiency R7081-100 10-inch PMTs from Hamamatsu will be used as part of the dichroicon assembly. Overall, the photocathode coverage of the inner detector volume will be approximately 40\%.

The PMT support structure (PSUP) is being constructed to allow flexibility for future upgrades with additional PMTs, additional dichroicons, and alternative photon detection technology, such as LAPPDs or SiPMs.  The detector is surrounded by a muon veto, which can also be used to identify entrance and exit points to validate muon track reconstruction in \eos.

CAEN V1730 digitizers will be used to record PMT waveforms at 500 MSa/s. Interpolating between digitization samples using information about the PMT pulse shape will allow for the extraction of photon times to sub-ns precision. The V1730 can record up to a 1-ms long waveform per acquisition trigger with zero deadtime, which will allow detection of time coincidence signals such as antineutrinos. The front end will be read out over optical fibers to a CAEN A3818 PCIe controller mounted in the DAQ server. It is controlled by data acquisition software built on top of CAEN's high level digitizer software libraries. 

Central analog sums of PMT pulses and single-hit “trigger primitives” will be done by a central summing board whose design will be similar to the MTC/A boards used by the Sudbury Neutrino Observatory (SNO)~\cite{sno_electronics}.  The analog sum will allow triggering either proportional to total recorded charge or to a total number of hits, but will also allow topological triggering by summing only subsets of hits -- for example, triggering only on hits on the \eos barrel PMTs, or on the long-wavelength dichroicon PMTs, etc.  

Triggering will be done by a custom central trigger board: the ``Penn Trigger Board'' (PTB). The PTB uses a Xilinx Zynq system-on-a-chip to implement various logical combinations of signals from the detector, veto, or calibration sources, and will fan out a global trigger signal and a clock to each V1730. The trigger signal will also be digitized on the 16th channel of each V1730 in order to ensure time synchronization between all channels.  The first version of the PTB was developed for the Long Baseline Neutrino Experiment (LBNE) 35-ton demonstrator~\cite{ZHANG2020162826}, and subsequent iterations have been used at the ProtoDUNE detector~\cite{DUNE:2020cqd} and the SBND experiment~\cite{doi:10.1146/annurev-nucl-101917-020949}.

The PMTs will be powered by CAEN A7030P HV power supply boards, which can provide up to 3 kV and 1 mA per channel.  A Slow Control and Monitoring system will provide real-time monitoring of the detector state, environmental conditions, DAQ performance, and data quality in a centralized web-based interface.

Future upgrades to the \eos readout are being explored.  These include: development of custom front end electronics with triggerless continuous waveform digitization and faster sampling rates;
or using analog processing to extract features from the analog waveforms from the PMTs, before digitization is done.

Several deployed radioactive and optical sources will be used to characterize the event reconstruction capabilities and the optical properties of the detector components.  A light injection system (LIS) will supply specific wavelengths of light via a laser to a number of optical fibers mounted near the PMTs and directed at the opposite side of the detector.  This enables wavelength-dependent studies of photon scattering and absorption in the target liquid.  Radioactive sources and a light-diffusing source will be deployed vertically along the central axis of the cylindrical detector with the ability to control the azimuthal position of a source, facilitating the study of event position, direction, and energy reconstruction capabilities, as well as the calibration of detector efficiency and timing characteristics.  In particular, a directional source is being developed to emit a narrow stream of electrons that will be used to demonstrate the direction reconstruction capabilities.  Off-axis deployment ports have been incorporated into the OV lid to allow for sources to be deployed off-axis.  Plans are also being made to deploy a neutron-gamma source (such as $^{241}$Am-$^{9}$Be) on the outside of the OV for evaluating the direction reconstruction without optical interference from source deployment hardware.


\section{Analysis tools}

The \eos detector is modeled using the RAT-PAC simulation and analysis package~\cite{ratpac}.  This tool is based on the RAT package, which was initially developed for the Braidwood reactor experiment. The framework has since been adopted and developed by a number of neutrino and dark matter projects, such as SNO+~\cite{SNO:2015wyx}, MiniCLEAN~\cite{Wang:2017jts}, and \theia~\cite{Askins:2019oqj}. RAT-PAC is a Geant4-based~\cite{GEANT4:2002zbu} tool that incorporates a custom GLG4Sim-based scintillation model to provide improved modeling of optical photon generation, along with a procedural geometry description, database support, and full modeling of all stages of event simulation.  This includes: the initiating physics interaction(s); optical photon generation and propagation;  photon detection at the single PE level, including individual photon detector charge and timing response; and data acquisition with fully customizable simulation of front end electronics, trigger systems, and event builders. 

Material and optical properties of the potential scintillator target materials are based on bench-top measurements~\cite{CHESS2016,CHESS2020,Caravaca:2016fjg,Kaptanoglu:2021prv,Onken:2020pnv}, and photon detector properties are based on the properties reported by the manufacturer, Hamamatsu~\cite{ham_datasheet_r14688,ham_datasheet_r7081}, which will be updated based on a detailed ex-situ characterization of each individual PMT.

The event reconstruction capabilities of \eos are explored using both traditional likelihood and machine learning approaches~\cite{SNO:2011hxd,Eller:2022xvi}.  This allows for a complete understanding of the impact of material properties, photon detector response, and analysis methodology on the resulting performance.  Detailed performance predictions will be discussed in a future communication.

\section{Deployment plan}

\eos will initially be constructed at the University of California, Berkeley in the USA.  Detector design and large-scale procurement will proceed in 2022, construction in 2023, and commissioning, data taking and analysis are planned for 2024.

\eos will initially be filled with ultra-pure water, a well-understood target material to be used for the purposes of detector commissioning, and to baseline the event reconstruction capabilities and comparisons to MC model predictions.  Following this, increasing amounts of scintillator will be injected in order to achieve a WbLS cocktail, according to developments at BNL.  The full program of planned source deployments described below will be repeated at a range of scintillator concentrations, in order to fully explore the phase space of differing Cherenkov-to-scintillation ratios.  The detector has been designed to also allow for a pure LS fill, and for the potential to explore isotopic loading.

Calibrations will be performed regularly and as needed.  The relative timing of the photon detectors will be calibrated with the deployed light-diffusing source and will be performed depending on the observed stability of the PMTs and other detector conditions.  The stability of the various WbLS cocktails will be determined by regularly evaluating the attenuation of the target liquid using the LIS, along with calibrating the energy scale (light yield) using deployed radioactive sources.  The performance of the event reconstruction algorithms will be quantified as needed using radioactive sources, with additional deployments when any conditions change.  These studies will also make full use of the natural radioactivity present in the detector, which is useful for extracting the time profile of scintillating target liquids.  The multitude of particle species available from all of the expected sources (alpha, beta, gamma, neutron) will allow a comprehensive assessment of particle ID capabilities \cite{Ford:2022wla}.  

After the conclusion of operations at Berkeley, \eos could be repurposed for additional measurements.  This would take place after the end of the primary project period, and successful completion of the project objective. Options include:
\begin{itemize}
\item Deployment at a nuclear reactor for low-energy antineutrino reconstruction.  This provides a near-field demonstration of the remote reactor monitoring concept, and the first use of hybrid detector technology for low-energy antineutrino reconstruction with a fast scintillator.
\item Deployment at a particle test beam such as CERN or FNAL, for hadronic event reconstruction.  This would be useful for advanced event identification, including sub-Cherenkov threshold particle identification critical to a long baseline program.
\item Deployment at the SNS at ORNL for reconstruction of neutrinos in an energy regime relevant for supernova studies.  This could allow measurements of neutrino charged-current cross sections on oxygen and, in a potential enhanced configuration, lithium.  
Additional opportunities at this site include neutron studies, which would provide useful background characterization for antineutrino and rare event searches, as well as searches for physics Beyond the Standard Model (BSM physics).
\item Underground deployment.  This would allow more precise measurements of event reconstruction capability and background rejection in a low-background environment.
\end{itemize}

\section{Conclusion}
Hybrid neutrino detectors have the potential to revolutionize the field of low-- and high-energy neutrino detection, offering unprecedented event imaging capabilities and equivalent high background rejection.  These performance characteristics would substantially increase sensitivity to a broad program of fundamental and applied physics.

\eos will provide a flexible testbed for novel detector technology, and a validated model of hybrid detector technology that can be made available for use by the community.  The detector will be capable of demonstrating full event reconstruction (vertex, direction and energy) for events from MeV-scale to tens of MeV, with the ability to deploy both pure LS and WbLS at a range of concentrations. The flexible PSUP allows deployment of a range of sizes and types of photon detectors, including the ability to add dichroicons.

 
\section{Acknowledgements}

Work conducted at Lawrence Berkeley National Laboratory was performed under the auspices of the U.S. Department of Energy under Contract DE-AC02-05CH11231. The work conducted at Brookhaven National Laboratory was supported by the U.S. Department of Energy under contract DE-AC02-98CH10886. 
The project was funded by the U.S. Department of Energy, National Nuclear Security Administration, Office of Defense Nuclear Nonproliferation  Research and Development (DNN R\&D).
This material is based upon work supported by the U.S. Department of Energy, Office of Science, Office of High Energy Physics, under Award Number  DE-SC0018974.

\bibliographystyle{ieeetr}
\bibliography{bibliography.bib}

\end{document}